# Spectroscopic characterization of atmospheric pressure μ-jet plasma source

**N.Bibinov[1], N.Knake[2], H.Bahre[2], P.Awakowicz[1], V.Schulz-von der Gathen[2]**

[1] Institute for Electrical Engineering and Plasma Technology, Ruhr-Universität Bochum, Universitätsstr. 150, 44801 Bochum, Germany

[2] Institute for Applied Plasma Physics, Ruhr-Universität Bochum, Universitätsstr. 150, 44801 Bochum, Germany

Email: _Nikita.Bibinov@rub.de_, _svdg@ep2.rub.de_

**Abstract**. A radio frequency μ-jet plasma source is studied using He/$O_2$ mixture. This μ-jet can be used for different applications as a source of chemical active species e.g. oxygen atoms, molecular metastables and ozone. Using absolutely-calibrated optical emission spectroscopy and numerical simulation, the gas temperature in active plasma region and plasma parameters (electron density and electron distribution function) are determined. Concentrations of oxygen atoms and ozone in the plasma channel and in the effluent of the plasma source are measured using emission and absorption spectroscopy. To interpret the measured spatial distributions, the steady-state species' concentrations are calculated using determined plasma parameters and gas temperature. At that the influence of the surface processes and gas flow regime on the loss of the active species in the plasma source are discussed. The measured spatial distributions of oxygen atom and ozone densities are compared with the simulated ones.

PACS: 52.50.Dg, 52.80.Pi, 52.70.Kz

## 1. Introduction

The micro-scaled atmospheric pressure plasma jet (μ-APPJ) [1] operated under a flow of He/O$_2$ gas mixture and at an excitation frequency of 13.56 MHz is a "cold" source of reactive species such as atomic oxygen, molecular metastable and ozone. Despite the wide field of possible applications such as in bio-medicine, the plasma-chemical processes of these kind of jet-like plasma sources is not yet clear and must be studied for the purpose of optimization of the sources. High collision rates of atoms and molecules, small diameters of the plasma channel and their influence on heterogenic processes, and different wall materials beside the plasma channel are examples of peculiarities of this plasma source which complicate understanding of plasma-chemical processes and thereby the optimization processes itself.

To gain knowledge on the processes occurring in the plasma, both measurements and modelling have to be applied while the synergetic combination of both may ideally cancel out the limitations of each individual method. Since probe measurements of, e.g. inside the μ-APPJ are difficult due to the small plasma volumes, most of the experimental data are derived from spectroscopic measurements. Phase-resolved optical emission spectroscopy (PROES) [2] provides information on excitation of species as well as on the plasma parameters in the plasma channel with temporal and spatial resolution. This technique highly relies on sophisticated collisional radiative models and is often limited to relative data of emission ratios such as in actinometry [3]. Two-photon absorption laser-induced fluorescence spectroscopy (TALIF) allows quite reliable determination of species' densities but is quite demanding in terms of equipment and usually limited to one distinct species [4,5].

Apart from the challenge of covering multiple time scales (electron dynamics in the range of ns and chemistry in the range of ms), numerical simulations, on the other hand, give a convenient insight into almost all plasma parameters and densities but are highly depending on the choice of assumptions, species, reaction rates, cross-sections, boundary conditions, etc. thus leaving a huge number of free parameters which must be determined by experiments or in other words be benchmarked [6].

To study the processes in the plasma source, we apply a combination of two methods namely optical emission spectroscopy (OES) and numerical simulation which supplement each other and allow to solve the problems which arise when these two methods are used separately. The gas temperature in plasma source is determined by OES using the rotational distribution of diatomic molecules and numerical simulation of emission spectra. Plasma parameters (electron density and electron velocity distribution function) are determined applying numerical solution of Boltzmann equation and absolutely calibrated optical emission spectroscopy. To check the validity of determined plasma parameters and applied plasma-chemical model, we compare the simulated and measured densities of oxygen atoms and ozone in active plasma volume and in the effluent of the plasma source. The oxygen atoms and ozone densities in the plasma channel and in the effluent are measured applying emission and absorption spectroscopy. The measured species densities are compared with steady-state densities of oxygen atoms and ozone calculated by simulation of chemical kinetics based on plasma parameters determined for the studied experimental conditions.

## 2. Experimental setup

### 2.1 OES measurements

The μ-APPJ is a capacitively-coupled RF-plasma source (see figure 1) operated at an excitation frequency of 13.56 MHz. Two plane electrodes with thickness of 1 mm and length of 40 mm are separated by a gap of 1.3 mm (part A in figure 1). To exclude the influence of surrounding air on chemical kinetics in the effluent region and on the plasma conditions in the plasma channel, the μ-jet is supplied with non-conductive plates (of PVC) with the same thickness and length as the electrodes (part B in figure 1) and quartz windows on both sides. This μ-jet plasma source is operated in a He/$O_2$ (1500 sccm/ 22.5 sccm) mixture and at a RF-sender power of 30 W. Gas velocity in plasma and effluent channels amounts to 19.5 m/s.

We characterize the plasma conditions of μ-APPJ by applying OES and numerical simulation using the emission bands of nitrogen which have a well-known mechanism of excitation, and high intensities because of high probability of spontaneous emission. To prevent influence of nitrogen admixture on plasma conditions, the quantity of nitrogen admixture to He/$O_2$ gas mixture must be low. We use pure He (purity 5.0) and technical oxygen (purity 2.5) containing low admixture (about 0.5 %) of nitrogen. Concentration of nitrogen in total gas flow is not known and is determined. The emission spectrum of µ-jet plasma source is measured with a spatial resolution of about 1 mm using a broad-band echelle-spectrometer (ESA-3000) and gratings spectrometer QE65000 (Ocean Optics). The former measures the emission spectrum simultaneously in about 100 spectral orders in the spectral range of $\lambda = 200\text{-}800$ nm with a spectral resolution of 0.015-0.06 nm while the letter measures spectrum in the spectral range of $\lambda = 200\text{-}900$ nm with a spectral resolution of about 1.3 nm. Both spectrometers are relatively and absolutely-calibrated using wolfram ribbon lamp and well known UV-emission spectra of $N_2$ and NO [7]. To increase the spatial resolution of OES characterization, the optical fibers of the spectrometers are provided with a diaphragm that limits the acceptance angle to 1.4°.

Absorption spectroscopy is used for determination of ozone concentration produced in the µ-APPJ. For this, the spectrum of a deuterium lamp (HAMAMATSU L625) transmitted through the gas gap is measured using the grating spectrometer QE65000 (see figure 1). The thickness of the absorbing layer of 1 mm is determined by the quartz panes. During absorption and OES measurements, the optical fibers are shifted parallel to the electrodes' surface along the gas gap in the $x$-direction with a step of 2 mm.

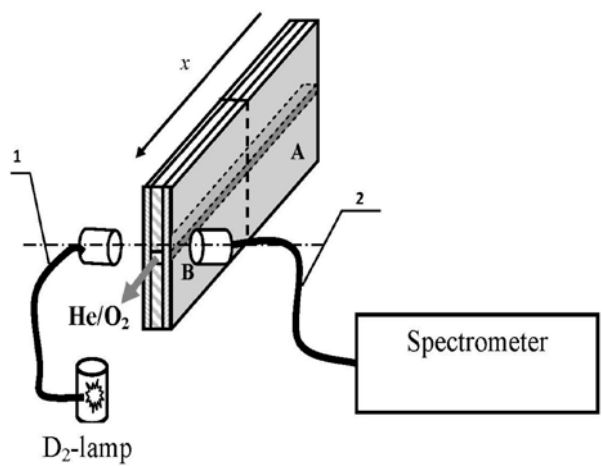

**Figure 1.** Experimental setup for *in situ* absorption measurements and OES characterization of µ-APPJ (13.56 MHz) operated in $He/O_2$ mixture at atmospheric pressure. Plasma source consists of part A with RF electrodes, and part B, where electrodes are replaced by plates of PVC. Gap between RF-electrodes (and PVC plates) amounts to 1.3 mm. One electrode is grounded. Active plasma volume and effluent are protected from the surrounding atmosphere by quartz panes. 1 – optical fiber of deuterium lamp, 2- optical fiber of spectrometer (ESA-3000 or QE65000) with diaphragm to confine acceptance angle.

The measured emission spectrum of μ-APPJ shows lines of helium and oxygen atoms ($O(^5S^o-^5P)$, $O(^3S^o-^3P)$), molecular bands of nitrogen molecules ($N_2$(C-B) and $N_2^+$(B-X)) and hydroxyl radical (OH(A-X)). The "second positive system" ($N_2$(C-B,0-0), λ=337.1 nm) and "first negative system" ($N_2^+$(B-X,0-0), λ=391.4 nm) of molecular nitrogen are detected in UV-spectral range and used for determination of plasma parameters and gas temperature.

The measured absorption spectrum (see figure 2) agrees with the well-known spectrum of ozone [8] in the spectral range 225 nm < λ < 300 nm. In the spectral range 200 nm < λ < 225 nm, the accuracy of absorption measurement decreases because of the reduced sensitivity of the applied spectrometer in this range.

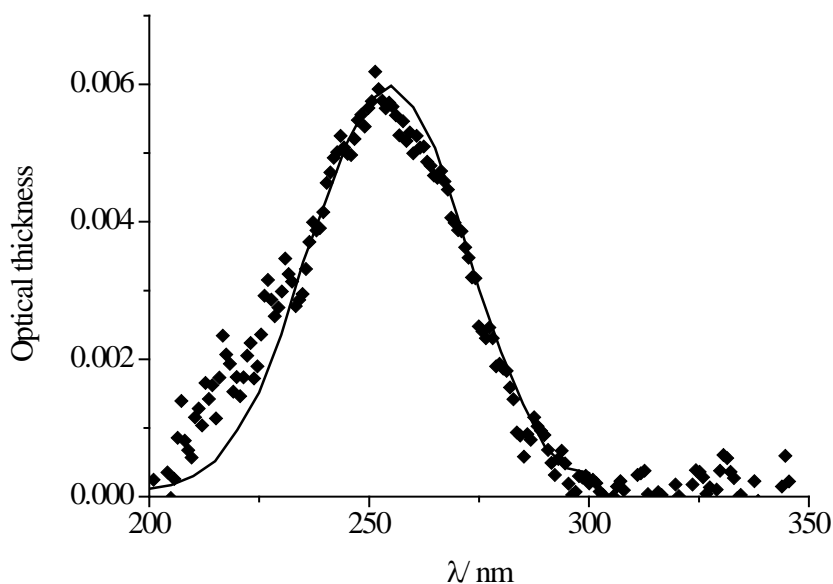

**Figure 2** . UV-absorption spectrum measured in the active plasma volume of the μ-APPJ (♦). Normalized absorption spectrum of ozone [8] (solid line). At λ = 253.7 nm, absorption coefficient of ozone is (3.072 ± 0.032)·$10^4$ $m^{-1}$ .

*2.3 Determination of gas temperature*

Gas temperature in active plasma volume is one of the most important parameter for simulation of plasma-chemical kinetics of plasma jet. The rate constants of chemical reactions and concentration of species depend on the gas temperature. We determine this parameter using nitrogen molecular emission in assumption that rotational and translation degrees of freedom of diatomic molecules have equal temperature at atmospheric pressure conditions. Rotational distribution in $N_2$(C-B,0-0, $\lambda$=337.1 nm) vibrational band is used for determination of the gas temperature ($T_g$) in the active plasma volume. We simulate this vibrational band with the same spectral resolution of echelle spectrometer for varied rotational temperatures, and compare simulated and measured spectra for determination of gas temperature (see figure 3).

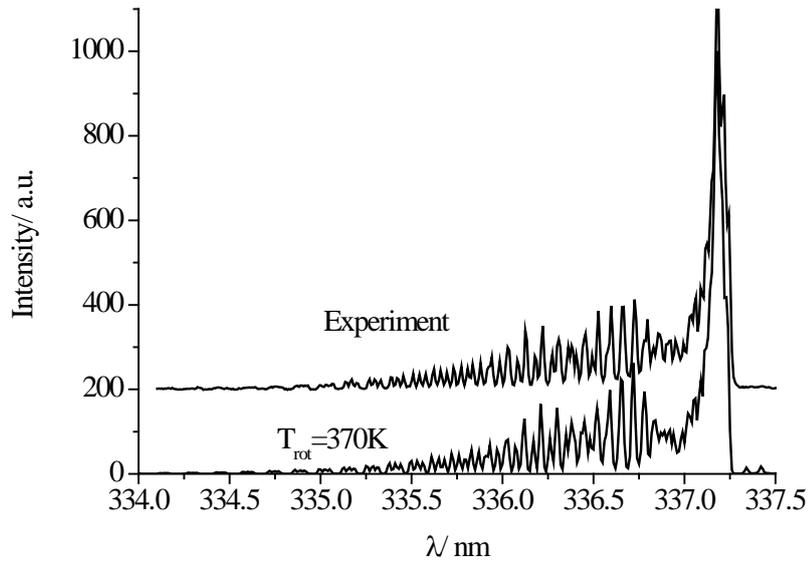

**Figure 3.** Rotational structure of $N_2$(C-B,0-0) vibrational band measured at the center of the plasma channel of μ-APPJ source operated with He/$O_2$ (1500 sccm/ 22.5 sccm) mixture. Simulated spectrum with rotational temperature of 370 K (bottom). Confidence interval amounts to ±20 K. The measured spectrum (top) is shifted for clarity.

Averaged rotational temperature with confidence interval in observed plasma volume is determined by fitting procedure. The gas temperature is approximately constant along plasma channel and amounts to 370 ± 20 K.

## 3. Simulations

### 3.1 Chemical kinetics model

Plasmas containing nitrogen and oxygen produce a large quantity of active species, and it is impossible to detect and quantify all these species. We approach this problem by simulation of chemical kinetics based on plasma parameters. Chemical reactions in $He/O_2/N_2$ mixture included in the model are listed in table 1 and table 2. In the simulation presented here, we include chemical reactions concerning the production and destruction of metastable molecules and atoms of oxygen and nitrogen, ozone and nitric oxide. Other reactions which are not included in this model have small rate constant or one (or both) of the reactants has too small concentration in our plasma conditions. We simulate chemical kinetics in μ-APPJ at steady-state conditions and determine densities of active species. The temperature dependencies of rate constants of some chemical reactions (No.10-20 in table 1 and No.29-33 in table 2) are presented. For other reactions, the Arrhenius temperature dependence is assumed.

The rate constants for recombination reactions of oxygen atoms with the assistance of oxygen molecules (No 14,16,18 in table 1) [9] multiplied by the factor of 0.54 [12] are used as the rate constants for recombination processes (No 15,17,19 in table 1) with the assistance of helium atoms.

Because of the small cross-section of the plasma channel - a rectangle with sides 1mm x 1.3 mm, the excited and chemical active species can be effectively lost on the surface of the walls despite high gas pressure, and therefore low diffusion velocity. Diffusion coefficients of $7.9 \cdot 10^{-5}$ m$^2$/s, $1.1 \cdot 10^{-4}$ m$^2$/s, $1.53 \cdot 10^{-4}$ m$^2$/s and $8.3 \cdot 10^{-5}$ m$^2$/s for diffusion of $O_2$, $N_2$, O and $O_3$ in He respectively [18,19] are used. For the simulation of diffusion of nitrogen atoms, the coefficient for diffusion of oxygen atoms is assumed because the difference of collision diameters of these atoms is smaller than 5% [20] .

If the plasma parameters (electron density and electron distribution function) are known, then the averaged excitation and dissociation rates of oxygen and nitrogen molecules (No.1-4 in table 1, No.26,27 in table 2) can be calculated for a spatial resolution of 1 mm. Determination of these plasma parameters by absolute calibrated OES and simulation is described in the following.

The temporal behavior of the chemically-active species produced in the discharge is described by the equation of continuity (1). Gain and loss of the active species by different reactions in the discharge channel and their diffusion (according to Fick's law) are considered.

$$\frac{\partial N_M}{\partial t} = \gamma \frac{23.2 \cdot D(T) \cdot N_M}{d^2} + \sum k_{MN}(T) \cdot N_M \cdot N_N + \nabla_x N_M \cdot v \qquad (1)$$

where $N_M$ and $N_N$ (in m$^{-3}$) are the densities of species involved in the chemical reaction, $D(T)$ (in m$^2 \cdot$s$^{-1}$) is the diffusion coefficient, and $k_{MN}$ (in m$^3 \cdot$s$^{-1}$) is the rate constant of the corresponding chemical reaction. This equation describes the temporal evolution of species' density due to diffusion (first term on the right hand side of equation (1)), due to different chemical reactions (second term) and motion of species with gas flow (third term) with velocity $v$ (in m·s$^{-1}$). $\nabla_x[M]$ is the density gradient of species $M$ in axial direction. For simplification of calculation procedure, cylindrical symmetry of the plasma is assumed with diameter ($d$, in m) equal to averaged distance between electrodes. Chemical kinetics is simulated using equation (1) which is solved numerically for O$_2$(a$^1\Delta$), O$_2$(b$^1\Sigma$), O, O($^1$D), N$_2$(A$^3\Sigma_u^+$), N, NO and O$_3$ concentrations simultaneously until steady-state density of the active species is attained.

Atomic species can be lost to the walls of the plasma source by adsorption, heterogenic recombination, production of molecules and desorption from the surface. The rate of the surface recombination of atoms depends on densities of physisorption and chemisorption centers on the surface [21], occupancy of chemisorptions centers, mobility of the atoms on the surface, etc. Therefore, properties of materials of the plasma source [22], chemical active impurities in working gas and on the surface of the plasma source wall, the surface temperature [23] and treatment of the surface by fluxes of ions strongly influence efficiency of heterogenic recombination of atoms on the surface of the plasma source wall. On active centers that are occupied by some atomic species or impurities, ozone molecules can be dissociated. The loss of atomic species and ozone in reactions (2-4) to the walls of the plasma source is described using diffusion process to the wall of the plasma channel mentioned above, and probabilities of these reactions ($\gamma$) in equation (1) are varied from 0 up to 1 during simulation. These probabilities are unknown *a priori* because of influence of the uncertain considering conditions of the wall's surfaces.

$$O + wall \rightarrow 1/2\ O_2 \qquad (2)$$

$$O_3 + wall \rightarrow O_2 + O \qquad (3)$$

$$N + wall \rightarrow 1/2\ NO \qquad (4)$$

The chemisorption centers on the walls at our experimental conditions are occupied mainly by oxygen atoms because of low concentration of nitrogen and other impurities. Therefore, reactions for NO molecule production by adsorption of oxygen atoms, and $N_2$ molecules by adsorption of nitrogen atoms are not taken into account. Other possible processes like reactions with charged species, helium atoms, nitrogen atoms and different molecular radicals are not included in the model because of low density of these species.

**Table 1.** Chemical reaction scheme of ozone, oxygen molecules and oxygen atoms included in the model for simulation of chemical kinetics in the μ-APPJ and in the effluent.

| No. | Reaction | $k_{No}$ | $k_{No}$ ($T_g$=370K) | Ref. |
|---|---|---|---|---|
| 1 | $O_2(X) + e \rightarrow O_2(a^1\Delta) + e$ | Rate constant is calculated from plasma parameters | | |
| 2 | $O_2(X) + e \rightarrow O_2(b^1\Sigma) + e$ | Rate constant is calculated from plasma parameters | | |
| 3 | $O_2(X) + e \rightarrow 2O + e$ | Rate constant is calculated from plasma parameters | | |
| 4 | $O_2(X) + e \rightarrow O(^1D) + O + e$ | Rate constant is calculated from plasma parameters | | |
| 5 | $O(^1D) + O_3 \rightarrow O_2(a^1\Delta) + O_2$ | $1.5\cdot10^{-17}$ m$^3$s$^{-1}$ | $1.7\cdot10^{-17}$m$^3$s$^{-1}$ | [9] |
| 6 | $O(^1D) + O_3 \rightarrow O_2(b^1\Sigma) + O_2$ | $7.7\cdot10^{-18}$ m$^3$s$^{-1}$ | $8.6\cdot10^{-18}$ m$^3$s$^{-1}$ | [9] |
| 7 | $O(^1D) + O_3 \rightarrow 2O_2$ | $2.3\cdot10^{-17}$ m$^3$s$^{-1}$ | $2.6\cdot10^{-17}$ m$^3$s$^{-1}$ | [9] |
| 8 | $O(^1D) + O_3 \rightarrow O_2 + 2O$ | $1.2\cdot10^{-16}$ m$^3$s$^{-1}$ | $1.3\cdot10^{-16}$ m$^3$s$^{-1}$ | [9] |
| 9 | $O(^1D) + O \rightarrow O + O$ | $7.5\cdot10^{-17}$ m$^3$s$^{-1}$ | $8.3\cdot10^{-17}$ m$^3$s$^{-1}$ | [10] |
| 10 | $O(^1D) + O_2 \rightarrow O_2 + O$ | $6.4\cdot10^{-18}\exp(67/T)$ m$^3$s$^{-1}$ | $7.6\cdot10^{-18}$m$^3$s$^{-1}$ | [10] |
| 11 | $O + O_3 \rightarrow O_2(a^1\Delta) + O_2$ | $6.3\cdot10^{-18}\exp(-2300/T)$ m$^3$s$^{-1}$ | $1.3\cdot10^{-20}$ m$^3$s$^{-1}$ | [9,11] |
| 12 | $O + O_3 \rightarrow O_2(b^1\Sigma) + O_2$ | $3.2\cdot10^{-18}\exp(-2300/T)$ m$^3$s$^{-1}$ | $6.4\cdot10^{-21}$ m$^3$s$^{-1}$ | [9,11] |
| 13 | $O + O_3 \rightarrow 2O_2$ | $9.5\cdot10^{-18}\exp(-2300/T)$ m$^3$s$^{-1}$ | $2.0\cdot10^{-20}$ m$^3$s$^{-1}$ | [9,11] |
| 14 | $2O + O_2 \rightarrow O_2 + O_2$ | $1.9\cdot10^{-42}T^{-1}\exp(-170/T)$ m$^6$s$^{-1}$ | $6.8\cdot10^{-46}$ m$^6$s$^{-1}$ | [9] |
| 15 | $2O + He \rightarrow O_2 + He$ | $1.0\cdot10^{-42}T^{-1}\exp(-170/T)$ m$^6$s$^{-1}$ | $3.6\cdot10^{-46}$ m$^6$s$^{-1}$ | [9,12] |
| 16 | $2O + O_2 \rightarrow O_2(a^1\Delta) + O_2$ | $1.3\cdot10^{-42}T^{-1}\exp(-170/T)$ m$^6$s$^{-1}$ | $2.2\cdot10^{-45}$ m$^6$s$^{-1}$ | [9] |
| 17 | $2O + He \rightarrow O_2(a^1\Delta) + He$ | $7.0\cdot10^{-43}T^{-1}\exp(-170/T)$ m$^6$s$^{-1}$ | $1.2\cdot10^{-45}$ m$^6$s$^{-1}$ | [9,12] |
| 18 | $2O + O_2 \rightarrow O_2(b^1\Sigma) + O_2$ | $6.0\cdot10^{-43}T^{-1}\exp(-170/T)$ m$^6$s$^{-1}$ | $1.0\cdot10^{-45}$ m$^6$s$^{-1}$ | [9] |
| 19 | $2O + He \rightarrow O_2(b^1\Sigma) + He$ | $3.2\cdot10^{-43}T^{-1}\exp(-170/T)$ m$^6$s$^{-1}$ | $5.5\cdot10^{-46}$ m$^6$s$^{-1}$ | [9,12] |
| 20 | $O_2(a^1\Delta) + O_3 \rightarrow O + 2O_2$ | $5.0\cdot10^{-17}\exp(-2830/T)$ m$^3$s$^{-1}$ | $2.4\cdot10^{-20}$ m$^3$s$^{-1}$ | [11] |
| 21 | $O + 2O_2 \rightarrow O_3 + O_2$ | $5.5\cdot10^{-43}T^{-1.20}$ m$^6$s$^{-1}$ | $4.6\cdot10^{-46}$ m$^6$s$^{-1}$ | [13] |
| 22 | $O + O_2 + He \rightarrow O_3 + He$ | $3.3\cdot10^{-43}T^{-1.20}$ m$^6$s$^{-1}$ | $2.7\cdot10^{-46}$ m$^6$s$^{-1}$ | [13] |
| 23 | $O_2(a^1\Delta) + O \rightarrow O + O_2$ | $1.3\cdot10^{-22}$ m$^3$s$^{-1}$ | $1.4\cdot10^{-22}$ m$^3$s$^{-1}$ | [11] |
| 24 | $O_2(b^1\Sigma) + O_3 \rightarrow O + O_2(a^1\Delta)$ | $7.0\cdot10^{-18}$ m$^3$s$^{-1}$ | $7.8\cdot10^{-18}$ m$^3$s$^{-1}$ | [10] |
| 25 | $O_2(b^1\Sigma) + O_3 \rightarrow O + 2O_2$ | $1.5\cdot10^{-17}$ m$^3$s$^{-1}$ | $1.7\cdot10^{-17}$ m$^3$s$^{-1}$ | [10] |

**Table 2.** Chemical reaction scheme of nitrogen atoms and molecules and nitric oxide included in simulation of chemical kinetics in μ-APPJ and in the effluent.

| No. | Reaction | $k_{No}$ | $k_{No}$ (Tg=370K) | Ref. |
|-----|----------|----------|--------------------|------|
| 26 | $N_2(X) + e \rightarrow N_2(A^3\Sigma_u^+) + e$ | Rate constant is calculated from plasma parameters | | |
| 27 | $N_2(X) + e \rightarrow 2N + e$ | Rate constant is calculated from plasma parameters | | |
| 28 | $N + O + M \rightarrow NO + M$ | $1.8 \cdot 10^{-43} T^{-0.5}$ m$^6$s$^{-1}$ | $9.4 \cdot 10^{-45}$ m$^6$s$^{-1}$ | [11] |
| 29 | $2N + M \rightarrow N_2(X^1\Sigma_g^+, A^3\Sigma_u^+) + M$ | $8.3 \cdot 10^{-46} \exp(500/T)$ m$^6$s$^{-1}$ | $3.2 \cdot 10^{-45}$ m$^6$s$^{-1}$ | [11,14] |
| 30 | $N + O_2 \rightarrow NO + O$ | $1.1 \cdot 10^{-20} T \exp(-3150/T)$ m$^3$s$^{-1}$ | $8.2 \cdot 10^{-22}$ m$^3$s$^{-1}$ | [11] |
| 31 | $N_2(A^3\Sigma_u^+) + O_2 \rightarrow 2O + N_2$ | $1.6 \cdot 10^{-18}(T/300)^{0.55}$ m$^3$s$^{-1}$ | $1.8 \cdot 10^{-18}$ m$^3$s$^{-1}$ | [15] |
| 32 | $N + O_3 \rightarrow NO + O_2$ | $5.0 \cdot 10^{-18} \exp(-650/T)$ m$^3$s$^{-1}$ | $8.6 \cdot 10^{-19}$ m$^3$s$^{-1}$ | [11] |
| 33 | $O_3 + NO \rightarrow NO_2 + O_2$ | $9 \cdot 10^{-19} \exp(-1200/T)$ m$^3$s$^{-1}$ | $3.5 \cdot 10^{-20}$ m$^3$s$^{-1}$ | [11] |
| 34 | $2N_2(A^3\Sigma_u^+) \rightarrow N_2(A^3\Sigma_u^+) + N_2$ | $3.5 \cdot 10^{-16}$ m$^3$s$^{-1}$ | $3.9 \cdot 10^{-16}$ m$^3$s$^{-1}$ | [15] |
| 35 | $NO + N \rightarrow N_2 + O$ | $3.1 \cdot 10^{-17}$ m$^3$s$^{-1}$ | $3.4 \cdot 10^{-17}$ m$^3$s$^{-1}$ | [16] |
| 36 | $NO + O + M \rightarrow NO_2 + M$ | $9.1 \cdot 10^{-40} T^{-1.6}$ m$^6$s$^{-1}$ | $7.1 \cdot 10^{-44}$ m$^6$s$^{-1}$ | [17] |

where M = He, O$_2$

### 3.2 Determination of plasma parameters

To simulate chemical kinetics in μ-APPJ, excitation and dissociation rates (No.1-4 in table 1, and No.26,27 in table 2) ( $R^B$ in m$^3$s$^{-1}$) (5) of oxygen and nitrogen molecules are required

$$R^B = N_B \cdot n_e \cdot 4\pi \cdot \sqrt{2} \int_E f_v(E) \cdot \sqrt{\frac{2e}{m_e}} \cdot E \cdot \sigma^B(E) dE, \tag{5}$$

where $N_B$ - density of oxygen and nitrogen molecules in plasma volume (m$^{-3}$);

$n_e$ - electron density (m$^{-3}$) ;

$f_v(E)$ - electron velocity distribution function (EVDF) (eV$^{-3/2}$) normalized to fulfill the equation

$$4\pi \cdot \sqrt{2} \int_E f_v(E) \cdot \sqrt{E} dE = 1$$

$\sigma^B(E) = \sigma^B_{exc}(E)$ or $\sigma^B_{dis}(E)$ - cross section of electron impact excitation and dissociation of molecules correspondingly ($m^2$) [24,25];

$m_e$ , e - the electron mass (kg) and elementary charge (C);

$E$ – the electron kinetic energy (eV).

Plasma parameters ($n_e$ and $f_v(E)$) are required by calculation of reaction rates (5). For determination of plasma parameters at atmospheric pressure conditions, numerical simulation can be applied where Boltzmann equation for positive and negative ions and for electrons together with Poisson equation is solved. Many reactions of charged particles with not always well known cross sections must be taken into account and several assumptions have to be made. Therefore, excitation and dissociation rates of molecules calculated applying these plasma parameters can be unreliable. To enhance reliability of chemical kinetic simulation in μ-APPJ, we use a combination of two diagnostics namely numerical simulation and optical emission spectroscopy. We solve Boltzmann equation for electrons using variable electric field strength, simulate electron distribution function, calculate emission spectra for variable electric field and determine plasma parameters by comparison of measured and calculated spectra. By solution of Boltzmann equation, only loss of electron kinetic energy must be taken into account and therefore only most important cross sections of electron collisions with heavy particles are needed. These cross sections are well known for various atomic and molecular gases and hence method can provide more reliable plasma parameters and excitation rates.

### 3.2.1 Electron velocity distribution function

To determine the required plasma parameters, we compare measured and calculated emission spectra. The ratio of emission intensities of ionized molecular nitrogen and of neutral molecular nitrogen strongly depends on the RF-electric field ( $\vec{E}$ ) and is used for determination of electron velocity distribution function. To calculate the emission spectrum, we simulate the electron velocity distribution function in He/$O_2$ mixture at atmospheric pressure by numerical solution of Boltzmann equation in "local" approximation (6) using the program code "EEDF" developed by Napartovich *et al* [26,27].

$$\frac{\partial f}{\partial t} - \nabla_v(\frac{e}{m_e}\vec{E}f) = S_{coll} \quad , \qquad (6)$$

where $S_{coll}$ - collision integral for electrons with heavy species.

″Local approximation″ means that at every point in the plasma volume, EVDF can be determined from Boltzmann equation for homogeneous non-bounded plasma using local electric field strength. The "local approximation" is valid for highly collisional plasma such as the plasma at atmospheric pressure conditions [28]. Electric field strength in μ-APPJ is not known forehand. Hence, we simulate EVDF for different values of RF electric field. Using simulated EVDF, we calculate intensities of nitrogen molecular emissions and compare them with measured ones to determine the EVDF and also the electric field strength in active plasma volume.

**Table 3.** Elementary processes considered in simulations of EVDF in μ-APPJ operated in He/O$_2$ mixture at atmospheric pressure.

| No. | Reaction | No. | Reaction |
|---|---|---|---|
| 37 | O$_2$(X) + e → O$_2$(a$^1\Delta$) + e | 12 | He + e → He(2$^3$S, 19.83 eV) + e |
| 38 | O$_2$(X) + e → O$_2$(b$^1\Sigma$) + e | 13 | He + e → He(2$^1$S, 20.63 eV) + e |
| 39 | O$_2$(X) + e → 2O($^3$P) + e | 14 | He + e → He(2$^3$P, 20.96 eV) + e |
| 40 | O$_2$(X) + e → O($^3$P) + O($^1$D) + e | 15 | He + e → He(2$^1$P, 21.23 eV) + e |
| 41 | O$_2$(X) + e → O$_2$(X,R) + e | 16 | He + e → He(3$^3$S, 22.72 eV) + e |
| 42 | O$_2$(X,v″=0) + e → O$_2$(X,v″=1-4) + e | 17 | He + e → He(3$^1$P, 23.10 eV) + e |
| 43 | O$_2$(X) + e → O$_2$(4.5eV) + e | 18 | He + e → He(4$^1$P, 23.74 eV) + e |
| 44 | O$_2$(X) + e → O$_2^+$ + 2e | 19 | He + e → He(5$^1$P, 24.04 eV) + e |
| 45 | O$_2$(X) + e → O + O$^+$ + 2e | 20 | He + e → He$^+$ + 2e |
| 46 | O$_2$(X) + e → O + O$^-$ | | |
| 47 | O$_2$(X) + e → O$_2^-$ | | |

where O$_2$(X,R), O$_2$(X,v″=1-4) - rotational and vibrational excited oxygen molecules in ground electronic state respectively

Elementary processes considered for simulation of the electron distribution function are presented in table 3. The plasma electrons are accelerated in electric field and lose kinetic energy in elastic and inelastic collisions with helium atoms and oxygen molecules. Possible influence of nitrogen impurity in He/$O_2$ mixture on calculated electron distribution function is determined. This influence is very low and is neglected in simulation.

Intensities of the molecular vibrational bands $N_2$(C-B,0-0, $\lambda$=337.1 nm) and $N_2^+$(B-X,0-0, $\lambda$=391.4 nm) are measured using calibrated echelle spectrometer. Emission of molecular nitrogen ions can be excited by electron impact (7) and by Penning ionization of nitrogen molecules (8) through collisions with helium metastable atoms He($2^{1,3}$S,$2^3$P) excited by electron impact (9).

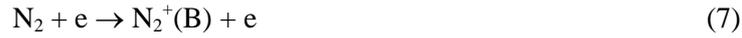

$$N_2 + e \rightarrow N_2^+(B) + e \qquad\qquad (7)$$

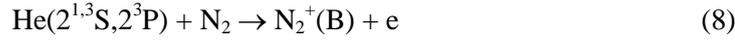

$$He(2^{1,3}S,2^3P) + N_2 \rightarrow N_2^+(B) + e \qquad\qquad (8)$$

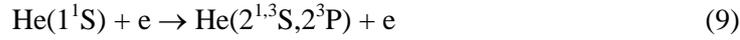

$$He(1^1S) + e \rightarrow He(2^{1,3}S,2^3P) + e \qquad\qquad (9)$$

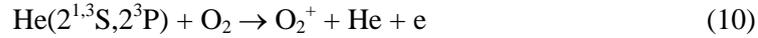

$$He(2^{1,3}S,2^3P) + O_2 \rightarrow O_2^+ + He + e \qquad\qquad (10)$$

At our plasma conditions, mainly metastable atoms He($2^{1,3}$S,$2^3$P) and resonance state He($2^1$P) are excited during electron impact. Rate constants for electron impact excitation of other excited states of helium atom are lower by more than two order of magnitude. Lifetime of resonance state He($2^1$P) concerning spontaneous emission amounts to 0.56 ns. The resonance photons ($\lambda$ = 584.3 nm) are "trapped" in helium and are absorbed by admixture molecules namely $O_2$ because of enough high concentration and high absorption coefficient [29]. Nitrogen and oxygen molecules are ionized by collisions with metastable helium atoms. Because of relative high concentration of $O_2$ and high rate constant [30], the Penning ionization of oxygen molecules (10) is the main quenching process of He($2^{1,3}$S,$2^3$P) metastable. Therefore, steady state concentration of helium metastables ($N_{He_{met}}$) is presented as:

$$N_{He_{met}} = \frac{N_{He} \cdot k_9 \cdot n_e}{k_{10} \cdot N_{O_2}} \qquad\qquad (11)$$

During collisions of helium metastables with nitrogen molecules, $N_2^+(X)$, $N_2^+(A)$ and $N_2^+(B)$ ions can be produced. A branching factor of 0.38 for excitation of $N_2^+(B)$ during Penning ionization of nitrogen molecule [31] is used in our model. By OES, emission transition between ground vibrational states of $N_2^+(B)$ and $N_2^+(X)$ is considered. Vibrational distribution of $N_2^+(B)$ excited in Penning ionization (8) corresponds to the Frank-Condon factors for $N_2(X) \rightarrow N_2^+(B)$ transition [31]. Accordingly to that, population of ground vibrational level of $N_2^+(B,0)$ amounts to 0.89 [32].

The steady state concentration of $N_2^+(B,0)$ at our plasma conditions can be determined using balance equation (12).

$$N_{N_2} \cdot n_e \cdot k_{exc}^{N_2^+(B,0)} + \frac{b \cdot N_{N_2} \cdot n_e \cdot N_{He} \cdot k_9 \cdot k_8}{N_{O_2} \cdot k_{10}} = N_{N_2^+(B,0)} \left( A_{N_2^+(B-X,0)} + k_q^{He} N_{He} + k_q^{O_2} N_{O_2} \right) \qquad (12)$$

Emission intensity is proportional to the steady state concentration of $N_2^+(B,0)$ molecules (13)

$$I_{N_2^+(B-X,0-0)} = \frac{A_{N_2^+(B-X,0-0)}}{A_{N_2^+(B-X,0)} + k_{N_2^+(B)}^{He} \cdot N_{He} + k_{N_2^+(B)}^{O_2} \cdot N_{O_2}} \cdot \left( N_{N_2} \cdot n_e \cdot k_{exc}^{N_2^+(B,0)} + \frac{b \cdot N_{N_2} \cdot n_e \cdot N_{He} \cdot k_9 \cdot k_8}{N_{O_2} \cdot k_{10}} \right) \qquad (13)$$

Intensity of neutral nitrogen emission $N_2(C-B,0-0)$ can be presented as (14)

$$I_{N_2(C-B,0-0)} = \frac{A_{N_2(C-B,0-0)}}{A_{N_2(C-B,0)} + k_{N_2(C)}^{He} \cdot N_{He} + k_{N_2(C)}^{O_2} \cdot N_{O_2}} \cdot \left( N_{N_2} \cdot n_e \cdot k_{exc}^{N_2(C,0)} \right) \qquad (14)$$

The ratio of intensities of nitrogen molecular bands (15) depends strongly on the electron distribution function .

$$\frac{I_{N_2(C-B,0-0)}}{I_{N_2^+(B-X,0-0)}} = \frac{\int_{333}^{338} I_\lambda d\lambda}{\int_{390}^{392} I_\lambda d\lambda} = \frac{N_{N_2} \cdot n_e \cdot Q_{N_2(C)} \cdot k_{exc}^{N_2(C,0)}}{N_{N_2} \cdot n_e \cdot Q_{N_2^+(B)} \left( k_{exc}^{N_2^+(B-X,0)} + \frac{b \cdot N_{He} \cdot k_9 \cdot k_8}{N_{O_2} \cdot k_{10}} \right)} \qquad (15)$$

where $\qquad Q_{N_2(C)} = \dfrac{A_{N_2(C-B,0-0)}}{A_{N_2(C-B,0)} + k^{O_2}_{N_2(C)} \cdot N_{O_2} + k^{He}_{N_2(C)} \cdot N_{He}} = 0.088$ and

$$Q_{N_2^+(B)} = \dfrac{A_{N_2^+(B-X,0-0)}}{A_{N_2^+(B-X,0)} + k^{O_2}_{N_2^+(B)} \cdot N_{O_2} + k^{He}_{N_2^+(B)} \cdot N_{He}} = 0.0506$$

are quenching factors for $N_2(C)$ und $N_2^+(B)$ excited states;

$A_{N_2(C-B,0)}, A_{N_2^+(B-X,0)}$ - the Einstein coefficients for spontaneous transitions ($s^{-1}$) [33, 34];

$k^{O_2}_{N_2(C)}, k^{O_2}_{N_2^+(B)}, k^{He}_{N_2(C)}, k^{He}_{N_2^+(B)}$ - the rate constants for quenching of respective nitrogen excited states by collisions with oxygen molecule and helium atoms ($m^3s^{-1}$) [33-36];

$k^{O_2}_{10} = 2.3 \cdot 10^{-16} \, m^3 s^{-1}$ and $k^{N_2}_8 = 7.1 \cdot 10^{-17} \, m^3 s^{-1}$ - the rate constants for Penning-ionization of oxygen and nitrogen molecules by collisions with helium metastables [30];

$b = 0.34$ – the branching ratio for excitation of $N_2^+(B,0)$ emission by Penning ionization (8) [30];

$k^M_{exc} = 4\pi \cdot \sqrt{2} \displaystyle\int_E f_v(E) \cdot \sqrt{\dfrac{2 \cdot e}{m_e}} E \cdot \sigma^M_{exc}(E) dE$ - the rate constant ($m^3s^{-1}$) for electron impact excitation of species $M$ (He,$N_2$) into the respective states;

$\sigma^M_{exc}(E)$ - the cross section of electron impact excitation for species $M$ (He, $N_2$) ($m^2$) into the respective states [24,37];

Quenching of excited nitrogen molecules by collision with other atomic and molecular species is not taken into account because of low densities of respective species in our experimental conditions.

Thus, we simulate electron velocity distribution function at variable RF-electric field, calculate intensities ratio of nitrogen molecular bands (15), compare the calculated ratio with a measured one to find the best fitted electric field and EVDF($x$) at variable distance $x$ from gas inlet of the plasma jet. Figure 4 shows electron distribution function at distance of 20 mm from the gas inlet of μ-APPJ operated with He/$O_2$ (0.985/0.015)) determined by this procedure. The RF electric field amounts to 310 V/mm.

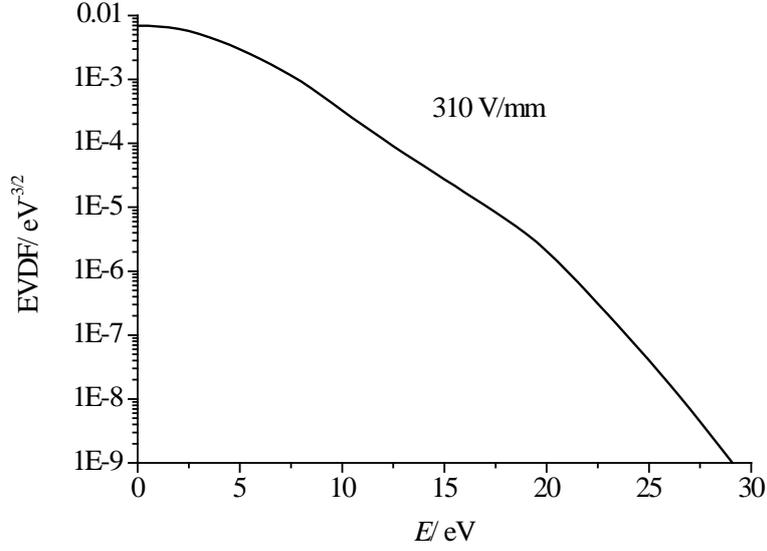

**Figure 4.** Electron velocity distribution function in the plasma channel 20 mm away from the gas inlet of μ-APPJ operated with He/O$_2$ (0.985/0.015) mixture determined using OES and numerical simulation.

### 3.2.2 Determination of electron density near the gas inlet ($n_e^0$)

In the second step, we determine another starting parameter for simulation of chemical kinetics namely, electron density using the oxygen photoemission. The atomic lines O($^3$S$^o$-$^3$P) at λ ~ 844.6 nm can be excited by electron-impact dissociative excitation (16) of oxygen molecule or by electron-impact excitation of oxygen atom in ground state (O(2p$^4$ $^3$P) ) (17)

$$O_2 + e \rightarrow O(2p^3 3p\,^3P) + O(2p^4\,^3P) + e \qquad (16)$$

$$O(2p^4\,^3P) + e \rightarrow O(2p^3 3p\,^3P) + e \qquad (17)$$

Lifetime of the excited state O(2p$^3$3p $^3$P) concerning spontaneous transition at λ~844.6 nm amounts to 33.3 ± 1.7 ns [32]. The rate constant of quenching of this excited state during collisions with oxygen molecules is equal to (7.8 ± 0.8)·10$^{-16}$ m$^3$s$^{-1}$ [38]. Excitation cross sections for processes (16) and (17) are presented in Ref. [33] and [40-42], correspondingly.

We determine a steady-state value of the electron density in the plasma channel using $O(^3S^o-^3P)$ emission near the μ-APPJ gas inlet ($x = 0$). In contrast to the other points in the plasma jet where oxygen atom density increases also due to transport of the atoms with gas flow from other parts of the jet, mechanism of atom production near the gas inlet includes only dissociation of oxygen molecules. We simulate plasma chemical kinetics based on the determined electron distribution function, and determine electron density at gas inlet, $n_e^0 = 6.0 \cdot 10^{16}$ m$^{-3}$ that corresponds to measured total intensity of $O(^3S^o-^3P)$ emission excited in processes (16,17).

### 3.2.3 Determination of spatial distribution of electron density in the plasma channel ($n_e$ ($x$))

The density of nitrogen molecule ($N_{N_2}^0$) in the plasma channel of $1.6 \cdot 10^{21}$ m$^{-3}$ is determined (18) using measured intensity of neutral nitrogen (N$_2$(C-B,0-0)) near the gas inlet of μ-APPJ ($I_{N_2(C-B)}^0$) and the plasma parameters ($n_e^0, f_v^0(E)$) determined in this point.

$$N_{N_2}^0 = \frac{I_{N_2(C-B,0-0)}^0 \cdot Q_{N_2(C)}^{-1}}{n_e^0 \cdot 4\pi \cdot \sqrt{2} \int\limits_E f_v^0(E) \cdot \sqrt{\frac{2 \cdot e}{m_e}E} \cdot \sigma_{exc}^{N_2(C,0)}(E)dE} \qquad (18)$$

Because of high gas flow and low probabilities of dissociation and ionization of nitrogen molecules in the active plasma volume, nitrogen concentration is considered constant along the plasma channel in $x$-direction ($N_{N_2}(x) = N_{N_2}^0$). This allows the determination of electron density along the plasma channel $n_e(x)$ from equation (19) using the measured nitrogen molecular emission intensity N$_2$(C-B,0-0) and the electron distribution function ($f_v^x(E)$) determined by the procedure described in subsection 3.2.1.

$$n_e(x) = \frac{I_{N_2(C-B,0-0)}(x) \cdot Q_{N_2(C)}^{-1}}{N_{N_2}^0 \cdot 4\pi \cdot \sqrt{2} \int\limits_E f_v^x(E) \cdot \sqrt{\frac{2 \cdot e}{m_e}E} \cdot \sigma_{exc}^{N_2(C,0)}(E)dE} \qquad (19)$$

### 3.3 Determination of oxygen-atom density

Now, the oxygen-atom density can be determined from equation (20) using the intensity of the O($^3$S$^o$-$^3$P) transition and the plasma parameters ($n_e(x)$ and EVDF(x)) and the cross sections $\sigma_{exc}^{O_2}(E)$ and $\sigma_{exc}^{O}(E)$ for the excitation processes (16) and (17), respectively.

$$N_O = \frac{I_{O(844)} \cdot Q_{O(844)}^{-1} - n_e \cdot k_{exc}^{O_2} \cdot N_{O_2}}{n_e \cdot k_{exc}^{O}},\qquad(20)$$

where $k_{exc}^Y = 4\pi \cdot \sqrt{2} \int\limits_E f_v(E) \cdot \sqrt{\frac{2 \cdot e}{m_e}} E \cdot \sigma_{exc}^Y(E) dE$ are the rate constants for excitation of emission O($^3$S$^o$-$^3$P) at $\lambda \sim 844.6$ nm by electron impact of $Y$ = O or O$_2$; $Q_{O(844)}^{-1}$ - is the quenching factor for the O(3p $^3$P) excited state [38].

## 4. Results and discussion

The results presented here are shown for a helium flow of 1500 sccm with admixture of 1.5% oxygen and RF-power of 30 W. Plasma source is operated in stable α-mode [2]. At steady-state conditions, plasma fills the whole gap between the electrodes (part A in figure 1). The RF-electric field determined using OES and simulation increases slightly when moving along the gas gap from the gas inlet point, and rapidly rises near the exit of the plasma channel (see figure 5). The determined electron density (using equation (19)) is approximately constant along the plasma channel except in the region near the gas inlet (see figure 6). The negative ions produced in the He/O$_2$ discharge could be the reason for variation of the plasma parameters. The negative ions are trapped within the plasma bulk because of positive plasma potential. But they are transported with the gas-flow parallel to the walls and electrodes. Therefore, the density of negative ions may increase with distance from the gas inlet. When negative ion concentration becomes higher than electron density, balance mainly between positive and negative ions concentrations provides quasi-neutrality of plasma. Positive ions are produced during electron impact ionization. For this electrons with kinetic energy higher than ionization potentials of helium (24.59 eV) and oxygen (12.07 eV) are needed which in turn necessitates high electric field required for electron acceleration at plasma conditions. To obtain enough high ionization rate in electronegative plasma such as the plasma far away from the gas inlet, the small number of electrons at this point should have enough high energy which is provided by the evaluated electric field.

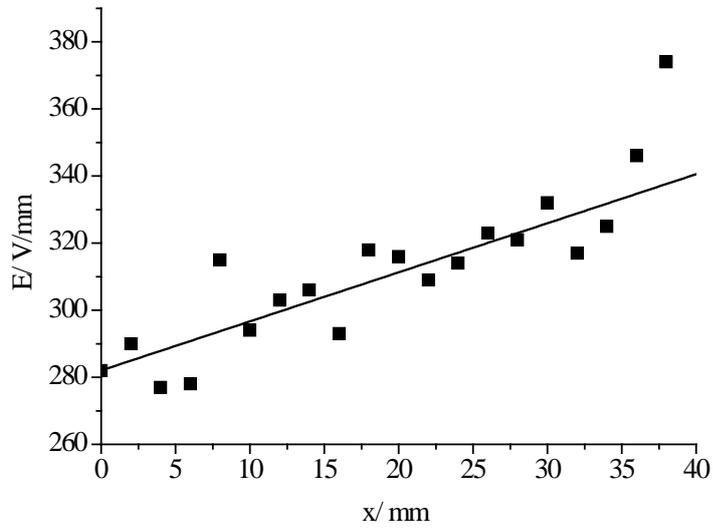

**Figure 5.** Spatial distribution of averaged RF-electric field determined applying OES and numerical simulation in μ-APPJ (He/$O_2$ mixture (1500 sccm/22.5 sccm), RF-power of 30 W).

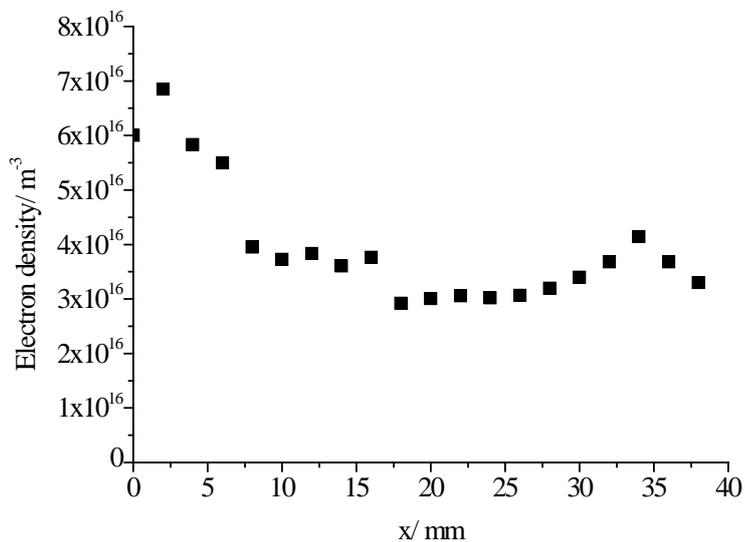

**Figure 6.** Spatial distribution of the electron densities in active plasma region of RF-plasma source determined using OES in μ-APPJ (He/$O_2$ mixture (1500 sccm/22.5 sccm), RF-power of 30 W).

By applying these plasma parameters and measured emission of $O(^3S^o-^3P)$ at $\lambda \sim 844.6$ nm, the steady-state oxygen atoms density is determined in plasma channel with spatial resolution (see figure 7). In the region up to 30 mm from the gas inlet, profile of oxygen density (presented in fig.7) corresponds to the data received in $He/O_2$ mixture applying TALIF measurements [4]. In TALIF measurements as well as in the experiments discussed here, the density of oxygen atom increases near the gas inlet und reach some saturation level. But the absolute values of oxygen density determined by OES is by a factor of about 3 higher than the same determined using TALIF. Furthermore, near the end of the plasma channel oxygen density measured in presented work decreases and deviates from the saturated value. Difference in experimental conditions namely, purity and concentration of applied oxygen and RF power between TALIF measurement and the considered experiment can cause deviation of species concentrations. The reason of these deviations will be studied in the future.

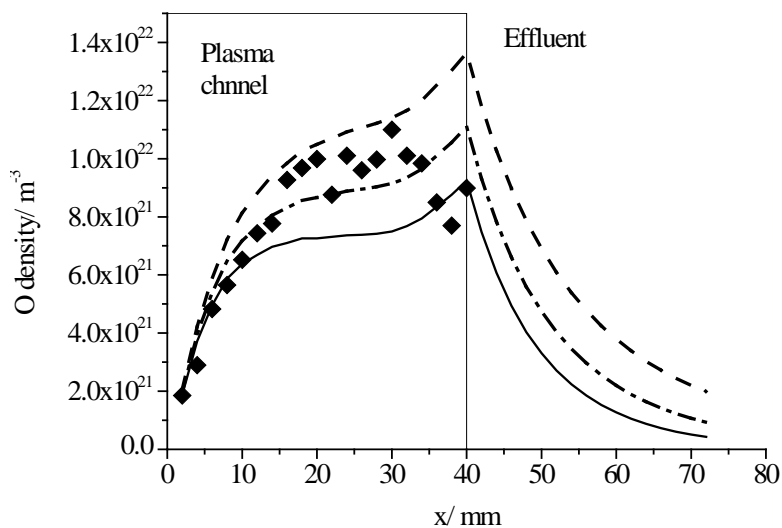

**Figure 7.** Spatial distribution of the steady-state oxygen atom density measured using OES in μ-APPJ ($\blacklozenge$) ($He/O_2$ mixture (1500 sccm/22.5 sccm), RF-power of 30 W). Simulated oxygen atom density for different efficiency of oxygen atoms' heterogenic recombination (dashed line - without heterogenic recombination of the atoms, dashed dot line - recombination probability of the oxygen atoms on the walls of 0.5, solid line - recombination probability of 1). The length of the active plasma channel amounts to 40 mm.

Ozone is produced in μ-APPJ together with atomic species and molecular metastables of oxygen. Steady-state concentrations of ozone in plasma channel and in the effluent as measured by applying absorption spectroscopy are shown in figure 8. The ozone density rises to a value of $4 \cdot 10^{21}$ m$^{-3}$ up to a distance of about $x = 20$ mm. Close to the nozzle ($x = 40$ mm), the density rises to a maximum value of $5 \cdot 10^{21}$ m$^{-3}$ and gradually falls off in the effluent region ($x = 40$ - $70$ mm), and approaches to the value of $2 \cdot 10^{21}$ m$^{-3}$ . This behavior is in contradiction to measurements with molecular beam mass spectrometry (MBMS) [43] in the effluent of μ-APPJ (0.6 % O$_2$ in He) "unprotected" from the surrounding atmosphere. A significant increase of ozone density with distance from the nozzle was found and maximum value of $1.5 \cdot 10^{21}$ m$^{-3}$ at a distance of 20 mm was measured. Probably, the observed differences can be explained as the influence of additional arrangement protected from surrounding atmosphere in our experiment and also the difference in oxygen admixture.

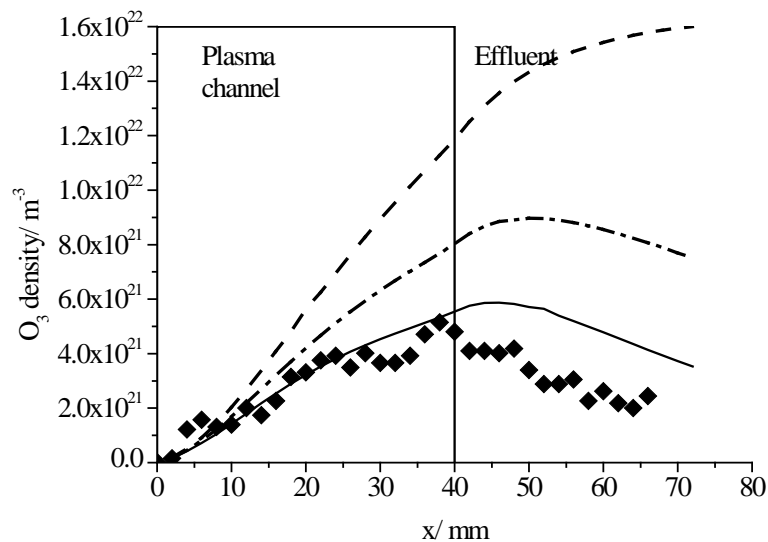

**Figure 8.** Ozone density measured using absorption spectroscopy (♦) in μ-APPJ (He/O$_2$ mixture (1500 sccm/22.5 sccm), RF-power of 30 W) and simulated applying the presented plasma-chemical model (dashed line - without heterogenic recombination of the oxygen atoms and destruction of ozone molecules; dashed dot line - probabilities of the oxygen atoms heterogenic recombination and ozone molecules destruction on the wall of 0.5; solid line – probabilities of the oxygen atoms heterogenic recombination and ozone molecules destruction on the wall of 1).

*4.1 Simulation of chemical kinetics*

To interpret the measured spatial profiles of species densities shown in figure 7 and figure 8, we simulate the chemical kinetics in the plasma channel and in the effluent of the plasma source. From the determined plasma parameters and known cross sections [24,25], the dissociation and excitation probabilities of nitrogen and oxygen molecules (see figure 9) are calculated. Chemical kinetics in the plasma channel and in the effluent are simulated using the reaction scheme presented in table 1 and table 2. The probabilities of dissociation and excitation of nitrogen and oxygen molecules are shown in figure 9. The data presented in this figure shows that oxygen atoms are dissociated in the plasma channel with approximately constant efficiency, and a moderate increase near the end of the plasma channel due to increase in electric field. Therefore, as shown in figure 8, ozone production will be increased by extending the plasma channel and at constant RF voltage, which is a requirement for discharge stability and for prevention of "arcing mode".

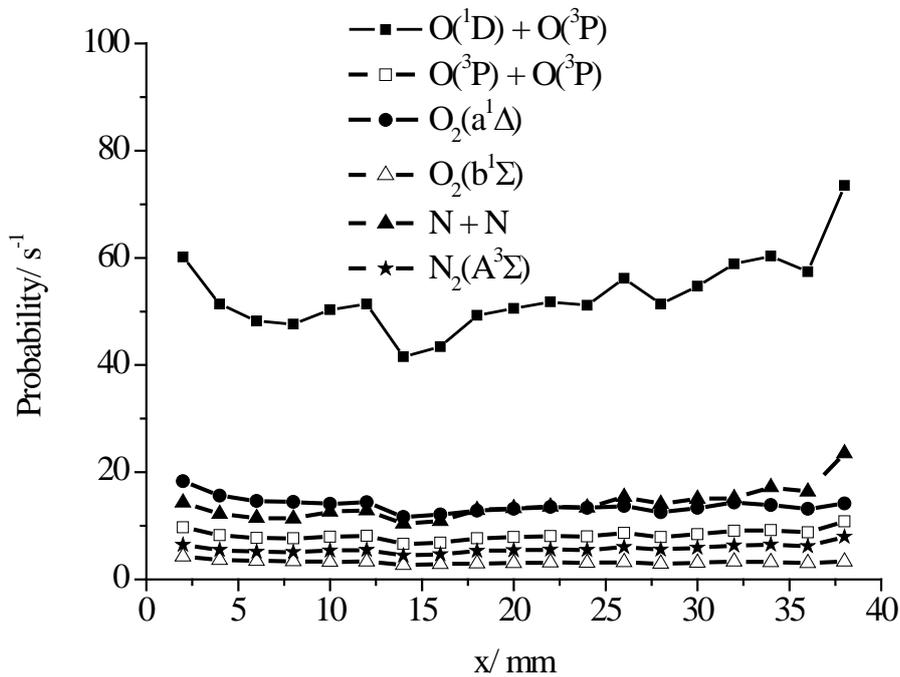

**Figure 9**. Spatial distribution of dissociation and excitation probabilities calculated in the frames of presented plasma-chemical model using measured plasma parameters.

The probabilities of heterogenic processes  (2-4) can be varied from 0 up to 1 depending on the surface temperature, morphology, chemical composition of the particles adsorbed on the surface of plasma source walls, etc. but are not known *a priori*.  So to compare calculated densities of oxygen atoms and ozone molecule with measured ones (see figure 7 and figure 8), we simulate chemical kinetics in the plasma channel and in the effluent with probabilities of heterogenic processes of 0, 0.5 and 1. The simulated oxygen densities (figure 7) for all probability values increase with distance from the  µ-APPJ gas inlet because of atoms transported by high gas flow. In the effluent region, the oxygen atoms density decreases gradually because of recombination processes (No.14-19 and 21,22 in the table 1).

The values of oxygen atom densities measured using OES and numerical simulation fall in between simulated ones (see figure 7). Recombination of the atoms on the µ-APPJ walls is important but not the main process by which the atoms are lost. By variation of the probabilities of heterogeneous processes from 0 to 1, the simulated steady-state concentration of oxygen atoms decreases by only 30 %.

The known values of recombination probability of oxygen atoms on the glass and metal surfaces amounts to about 0.01 and 0.2, respectively [21-23,44]. This is in accordance with the measured and simulated data in the region near the gas inlet (see figure 7). The oxygen atom density measured at distances for up to 30 mm from the gas inlet place between curves simulated for probabilities of heterogenic processes of 0 and 0.5.

One can see that, within the frame of the described model, the decrease of  oxygen atoms density near the end of the plasma channel cannot be simulated. The simulated curves show increase of oxygen density near the end of the plasma channel because of increase of electric field. But the measured oxygen density in this region decreases  the reason of which is not clear now and must be investigated through variation of plasma conditions.

The ozone density in the plasma channel and in the effluent of the plasma source is also simulated for the same plasma conditions and by variation of the probabilities of heterogenic processes (see figure 8). The measured ozone density agrees with simulated one in the active plasma volume ($x = 0$-40 mm) in assumption of probabilities of 1 for all heterogenic processes (2-4) that are explicitly overstated.  But also under this assumption, the ozone density in effluent region is about factor 2 lower than the simulated one.  Probably some process (or processes), which is responsible for ozone destruction in the plasma channel and in the effluent,  is not taken into account in the applied plasma-chemical model.

In the frame of the applied plasma-chemical model, the active species (atoms and molecular metastables) are produced during electron impact (No.1-4 in table 1 and No.26,27 in table 2). Only reactions with neutral species are included in the reaction scheme concerning production and  loss of oxygen, nitrogen atoms, oxygen molecular metastables, ozone and nitric oxide. Reactions between considered  species and charged particles (electrons, positive and negative ions) possess high rate constant [45] but densities of these species are not high enough to affect chemical kinetics. According to our estimations, the density of charged particles must be higher than $10^{18}$ $m^{-3}$ for appreciable influence on simulated densities of active species.  At our experimental conditions, electron density amounts to about $5 \cdot 10^{16}$ $m^{-3}$, and electronegativity (ratio of positive ion density to electron density)  is usually not higher than 10 in plasmas with oxygen. This means the densities of charged species are lower than $10^{18}$ $m^{-3}$ and their influence on chemical kinetics can be neglected.

Molecular nitrogen, which is determined as impurity in plasma channel, can in general affect  the production and loss of oxygen atoms and ozone molecules. We compare the loss of these species by participation of nitrogen atoms with total loss of species simulated in the frame of the applied model. Simulated nitrogen atom density reaches a maxima of about $10^{19}$ $m^{-3}$ near the end of the plasma channel. This steady-state concentration corresponds to lifetime of oxygen atoms and ozone molecule, considering loss by reaction with nitrogen atoms (No.29,33 in table 2), which amounts to about 532 ms and 116 ms, correspondingly. This lifetime are much longer than the  residence time of the species (about 2 ms) in plasma channel considering the gas flow. Therefore, influence of nitrogen impurity on the loss of oxygen atoms and ozone molecule in plasma source can be neglected.

The lifetime of nitrogen molecules concerning dissociation amounts to about 170 ms. This value is much longer than the residence time of nitrogen molecules in the plasma channel and, as it was assumed in the introduction of plasma chemical model of μ-APPJ (in section 3.2.3), the density of nitrogen molecule in the entire whole plasma channel is constant.

Different molecular and atomic species are produced in μ-APPJ using He/O$_2$/N$_2$ mixture. The steady state densities of these species in the plasma channel and in the effluent can be determined using experimental methods namely TALIF [4,5] and MBMS [43]. However, experimental detection of other species such as oxygen and nitrogen molecular metastables is complicated. This poblem can be solved by plasma chemical simulation using the determined plasma parameters and the known mechanisms of production and destruction of these species.

The metastables of oxygen molecules are excited in the μ-APPJ and flow along with the working gas. The steady state densities of these species are calculated within the frame of the model presented here. O$_2$(a$^1\Delta$, $E$=0.98 eV) density reaches a maximum of about 7.5·10$^{21}$ m$^{-3}$ near the end of the plasma channel (at x = 40 mm, see figure 10). In the effluent, the concentration of this metastable is high whereas the concentration of other oxygen molecular metastable namely O$_2$(b$^1\Sigma$, $E$=1.64 eV) is low because of effective quenching by collisions with ozone molecules.

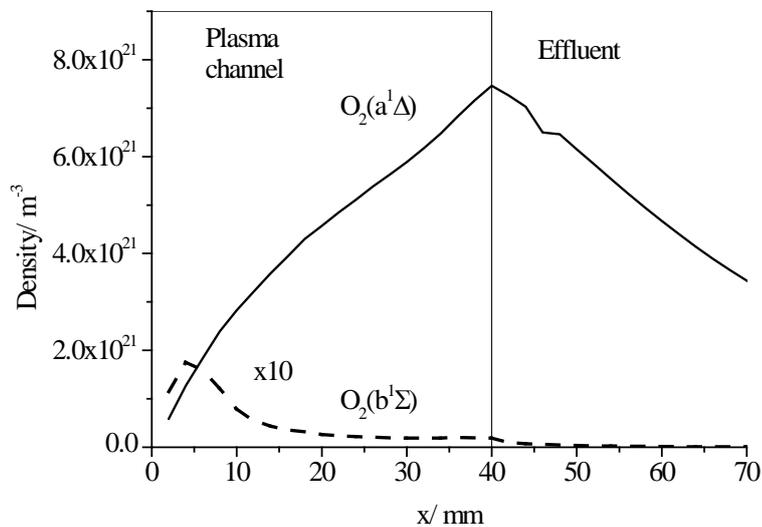

**Figure 10**. Densities of oxygen metastables calculated applying presented plasma-chemical model without heterogenic loss of active species (solid line - density of O$_2$(a$^1\Delta$), dashed line - O$_2$(b$^1\Sigma$) multiplied by 10).

## 5. Conclusion

By applying optical emission spectroscopy and numerical simulation, gas temperature and plasma parameters (electron density and electron velocity distribution function) in radio frequency atmospheric pressure μ-plasma jet operated in He/$O_2$ mixture are determined. The ozone and oxygen atom densities are measured using absorption and emission spectroscopy with spatial resolution of about 1 mm. Using determined plasma parameters, chemical kinetics in the plasma jet and in effluent are simulated, steady-state ozone and oxygen atom densities are calculated and compared with measured ones. Influence of heterogenic processes on chemical kinetics is discussed. According to our simulation, micro plasma jet operated in He/$O_2$ mixture can be used as source of atomic oxygen O($^3$P), oxygen molecular metastables $O_2$(a$^1\Delta$) and ozone with active species concentration of about 500 ppm.

**Acknowledgment**

Financial support by German Research Foundation of the projects A1 and C2 of Research Group FOR1123 "Physics of Microplasmas" is gratefully acknowledged.